\documentclass[twoside,11pt]{article}

\usepackage[b5paper, textwidth=130mm, textheight=190mm, centering]{geometry}

\usepackage{amssymb}
\usepackage[intlimits]{amsmath}
\usepackage{amsfonts}
\usepackage{amsthm}

\usepackage{graphicx}
\usepackage{mathpazo}
\usepackage{avant}

\usepackage[T1]{fontenc}
\usepackage{color}%T
\usepackage{listings}
\usepackage{algorithm}
\usepackage{algorithmicx}
\usepackage{algpseudocode}
\usepackage{cite}
\renewcommand\footnotemark{}

\usepackage{fancyhdr}
\begin{document}

%+Title
\title{Conditions for Unique Reconstruction of Sparse Signals Using Compressive Sensing Methods }

\author{Ljubi\v sa Stankovi\' c, Milo\v s Dakovi\' c, Srdjan Stankovi\' c, Irena Orovi\'c \thanks{The authors are with the University of Montenegro, 81000 Podgorica, Montenegro.}}

\date{}
\maketitle
%-Title

\thispagestyle{fancy}
\fancyhf{}
\fancyhead[CEO]{THE MONTENEGRIN ACADEMY OF SCIENCES AND ARTS PROCEEDINGS OF THE SECTION OF NATURAL SCIENCES, 22, 2017.}

%+Abstract
\begin{abstract}
    A signal is sparse in one of its representation domain if the number of nonzero coefficients in that domain is much  smaller than the total number of coefficients. Sparse signals can be reconstructed from a very reduced set of measurements/observations. The topic of this paper are conditions for the unique reconstruction of sparse
signals from reduced set of observations. After the basic definitions are introduced, the unique reconstruction conditions are reviewed using the spark, restricted isometry, and coherence of the measurement matrix.
Uniqueness of the reconstruction of signals sparse in the discrete  Fourier domain (DFT),
as the most important signal transformation domain, is considered as well.  
\end{abstract}
%-Abstract

\section{Introduction}

A discrete-time signal can be transformed into other domains  using different 
signal transformations. Some signals that cover the whole
considered interval in one domain could be sparse in a transformation domain,
i.e., could be located within a few nonzero coefficients. An observation or measurement is a linear combination of sparsity domain coefficients. Since the signal samples are linear combinations of the signal
transformation coefficients they could be considered as the observations of a
sparse signal in the transformation domain. Compressive sensing is a 
field dealing with a model for data acquisition including the problem of sparse 
signal recovery from a reduced set of observations \cite{donoho2006,Intro,Wiley,candes2006,YCE,5,bb11,MS_3,16a,INg,SS,MS_2,RobIAS}.  A reduced set of observations 
can be a result of a desire to sense a sparse signal with the lowest possible number of
measurements/observations (compressive sensing). It can also be a result of a
physical or measurement unavailability to take a complete set of observations \cite{Wiley}.
 In applications it could
happen that some arbitrarily positioned samples of a  signal are so heavily
corrupted by disturbances that it is better to omit them and consider as
unavailable in the analysis and to try to reconstruct the signal with a
reduced set of samples \cite{Impulsive,IMIMNCS,IMPSSL}. Although the reduced set of observations/samples
appears in the first case as a result of user strategy to compress the
information, while in the next two cases the reduced set of samples is not a
result of user intention, all of them can be considered within the unified
framework. Under some conditions, a full reconstruction of a sparse signal can
be performed with a reduced set of observations/samples, as in the case if a
complete set of samples/observations were available \cite{candes2006,RIP,RIPP,MP2,Uniq}. A priori information
about the nature of the analyzed signal, i.e., its sparsity in a known
transformation domain, must be used in this analysis. Sparsity is the main
requirement that should be satisfied in order to efficiently apply the
compressive sensing methods for sparse signal reconstruction.

Compressive sensing methods are successfully applied to many fields, including radar signal processing \cite{RadarCS,radarCS3,KnjgaTF,IRTR1,IRTR2,LJIM}, time-frequency analysis \cite{TFCS2,KnjgaTF,TF_sparse,ISAR22}, L-statistics \cite{LSSS,IMPSSL}, data hiding \cite{TFCS4}, communications \cite{CSBIOEL3}, image processing \cite{MS_4,Gradient_Isi}, etc.

Topic of this paper are conditions for the unique reconstruction of sparse signals from reduced set of observations/samples. The basic idea for unique reconstruction will be introduced through an illustrative and simple example in the next section. Then the unique reconstruction condition will be  explained within the spark, restricted isometry, and coherence framework. A special case of the signals sparse in the discrete  Fourier domain (DFT), as the most important signal transformation domain, will be considered at the end. A simple uniqueness criterion will be presented and illustrated on an example.   

\index{Sparse signals}%
\index{Compressive sensing}

\fancyhf{}
\fancyhead[LE,RO]{\thepage}
\fancyhead[RE]{\textit{Conditions for Unique Reconstruction}}
\fancyhead[LO]{\footnotesize \leftmark}

\section{Illustrative Examples}

Consider a large set of $N$ numbers $X(0)$, $X(1)$,...,$X(N-1)$. Assume that
only one of them is nonzero. We do not know either its position or its value. The aim is to find
the position and the value of this number. This case can be related to
many real life examples when we have to find one sample which differs from
other $N-1$ samples. The problem can easily be reformulated to the case when only one number differs from the expected and known value, and all other assume their expected-known values. 

The nonzero value  at an position $i$ will be denoted by $X(i)$. A direct way to find the position $i$ of nonzero
sample would be to perform up to $N$ measurements and compare each
$X(m)$ with zero. However, if $N$ is very large and
there is only one nonzero sample we can get the
result with just a few observations/measurements. A procedure for the reduced
number of observations/measurements is described next. 

Take random numbers as
weighting coefficients $a_{i}$, $i=0,1,2,...,N-1$, for each coefficient. Measure
the total value of all $N$ weighted coefficients, with weights $a_{i}$. Since only one of them is different from the known expected values $m_i$
(or from zero) we will get the total measured value%
\[
G=a_{1}m_1+a_{2}m_2+...+a_{i}(m_i+X(i))+...+a_{N}m_N.
\]
Next we will subtract the expected value $G_{T}%
=a_{1}m_1+a_{2}m_2+...+a_{N}m_N$ from $G$. The obtained
observation/measurement, denoted by $y(0)$, is
\[
y(0)=G-G_{T}=\sum_{k=0}^{N-1}a_{k}X(k)=a_{i}X(i),
\]
since the nonzero value in the space of $X(0)$, $X(1)$,...,$X(N-1)$ is
at one position only, $X(k)=X(i)\delta(k-i)$, $k=0,1,\dots,N-1$.

\begin{figure}[tb]
\begin{center}
\includegraphics{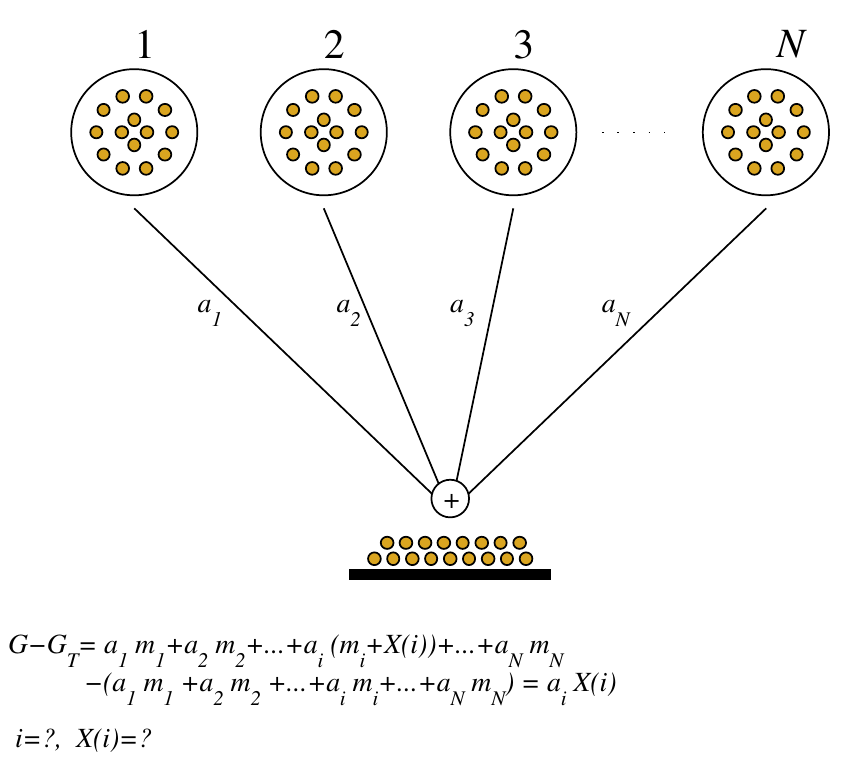}%
\caption{There are $N$ bags with coins. One of them, at an unknown position,
contains false coins. False coins differ from the true ones in mass for unknown
$X(i)=\Delta m$. The mass of the true coins in the $i$th bag is $m_i$.  }%
\label{kese_cs}%
\end{center}
\end{figure}

As an illustration consider a set of $N$ bags with coins. Assume that only one bag
contains false coins of a weight $m_i+X(i)$. It is different from the known
weights $m_i$ of true coins in bag $i$. The goal is to find the position and the difference
in weight of false coins. From each of $N$ bags we will take $a_{i}$,
$i=1,2,...N,$ coins, respectively. Number of coins taken from the $i$th bag is denoted by $a_{i}$.
The total measured weight of all coins from $N$ bags is $M$, Fig.\ref{kese_cs}%
. \ %

%TCIMACRO{\FRAME{ftbpFU}{4.5282in}{3.3486in}{0pt}{\Qcb{There are $N$ bags with
%coins. One of them, at an unknown position, contains false coins. False coins differ
%from the true ones in mass for unknown $X(i)=\Delta m$. The mass of the true
%coins is $m$. Set of coins for measurement is formed using $a_{1}$ coins from
%the first bag, $a_{2}$ coins from the second bag, an so on. The total measured
%value is $M=a_{1}m+...+a_{i}(m+X(i))+...+a_{N}m$. The difference of this value
%from the case if all coins were true is $M-M_{T}$. Equation for the case with
%one and two bags with false coins are presented (left and right).
%}}{\Qlb{kese_cs}}{kese_cs.eps}%
%width 4.5282in;  height 3.3486in;  depth 0pt;  original-width 4.078in;
%original-height 3.7418in;  cropleft "0";  croptop "1";  cropright "1";
%cropbottom "0";  filename 'Kese_CS.eps';file-properties "XNPEU";}}}%
%BeginExpansion
%EndExpansion
After the expected value is subtracted the observation/measurement $y(0)$ is
obtained%
\begin{equation}
y(0)=\sum_{k=0}^{N-1}X(k)\psi_{k}(0), \label{Hip0}%
\end{equation}
where the weighting coefficients for this measurement are denoted by $\psi
_{k}(0)=a_{k}$, $k=0,1,...,N-1$. In the space of unknowns (variables) $X(0)$,
$X(1)$,...,$X(N-1)$ this equation represents an $N$-dimensional hyperplane. We
know that only one unknown $X(k)$ is nonzero at an unknown position $k=i$.
The inter-section of hyperplane (\ref{Hip0}) with any of the coordinate axes could
be a solution of our problem. 

Assuming that a single $X(k)$ is nonzero, a solution will exist for
any $k$. Thus, one measurement would produce a set of $N$
possible single nonzero values equal to
\[
X(k)=y(0)/\psi_{k}(0)\text{, } \ \ \ \  \psi_{k}(0)\neq0\text{, }k=0,1,2,...,N-1.
\]
As expected, from one measurement we are not able to solve the problem and
to
find the position and the value of nonzero sample.
 
\begin{figure}[ptb]
\begin{center}
\includegraphics
{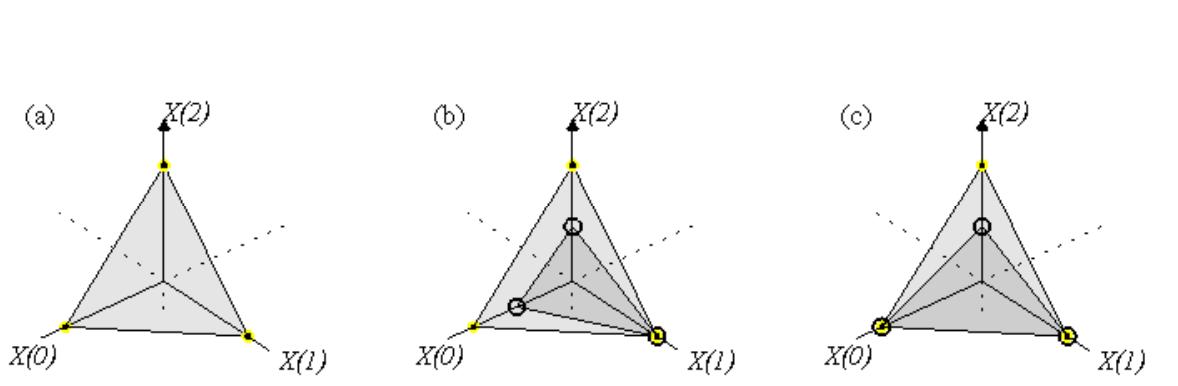}%
\caption{The solution illustration for $N=3,~ K=1,$ and various possible
cases: (a) Three possible solutions for one measurement plane. (b)  Unique
solution for two measurement planes. (c) Two possible solutions  for two
measurement planes.     }%
\label{projec_cs}%
\end{center}
\end{figure}

For $N=3$ possible solutions are illustrated with circles in Fig.\ref{projec_cs}a),
denoting intersections of measurements hyperplane with coordinate axes. 

%EndExpansion

If we perform one more measurement $y(1)$, with another set of weighting
coefficients $\psi_{k}(1)$, $k=0,1,...,N-1$, and get measured value
$y(1)=X(i)\psi_{i}(1)$ the result will be a hyperplane
\[
y(1)=\sum_{k=0}^{N-1}X(k)\psi_{k}(1).
\]
This measurement will produce a new set of possible solutions for each $X(k)$
defined by
\[
X(k)=y(1)/\psi_{k}(0),\text{ }k=0,1,2,...,N-1.
\]
If these two hyperplanes (sets of solutions) produce only one common value
\[
X(i)=y(0)/\psi_{i}(0)=y(1)/\psi_{i}(1).
\]
then it is the solution of our problem.

In a matrix form these two measurements can be written as
\begin{align*}
\left[
\begin{array}
[c]{c}%
y(0)\\
y(1)
\end{array}
\right]   &  =\left[
\begin{array}
[c]{cccc}%
\psi_{0}(0) & \psi_{1}(0) & ... & \psi_{N-1}(0)\\
\psi_{0}(1) & \psi_{1}(1) & ... & \psi_{N-1}(1)
\end{array}
\right]  \left[
\begin{array}
[c]{c}%
X(0)\\
X(1)\\
...\\
X(N-1)
\end{array}
\right] \\
\mathbf{y}  &  =\mathbf{AX}%
\end{align*}
where $\mathbf{A}$ is the matrix of coefficients (measurement matrix)
\[
\mathbf{A=}\left[
\begin{array}
[c]{cccc}%
\psi_{0}(0) & \psi_{1}(0) & ... & \psi_{N-1}(0)\\
\psi_{0}(1) & \psi_{1}(1) & ... & \psi_{N-1}(1)
\end{array}
\right]
\]
and $\mathbf{y}$ are observations/measurements of sparse variable $\mathbf{X}$.

Common value for two measurements $X(i)=y(0)/\psi_{i}(0)$ and $X(i)=y(1)/\psi
_{i}(1)$ is unique if
\[
\psi_{i}(0)\psi_{k}(1)-\psi_{i}(1)\psi_{k}(0)\neq0
\]
for any $i\neq k$.

In order to prove this statement assume that two different solutions $X(i)$
and $X(k),$ for the case of one nonzero coefficient, satisfy the same
measurement hyperplane equations%
\[
\psi_{i}(0)X(i)=y(0)\text{, \ \ }\psi_{i}(1)X(i)=y(1)
\]
and
\[
\psi_{k}(0)X(k)=y(0)\text{, \ \ \ }\psi_{k}(1)X(k)=y(1).
\]
Then
\begin{align*}
\psi_{i}(0)X(i)  &  =\psi_{k}(0)X(k)\\
&  \text{and}\\
\psi_{i}(1)X(i)  &  =\psi_{k}(1)X(k).
\end{align*}
If we divide these two equations we get
\[
\psi_{i}(0)/\psi_{i}(1)=\psi_{k}(0)/\psi_{k}(1)
\]
or $\psi_{i}(0)\psi_{k}(1)-\psi_{i}(1)\psi_{k}(0)=0$. This is contrary to the
assumption that $\psi_{i}(0)\psi_{k}(1)-\psi_{i}(1)\psi_{k}(0)\neq0$.

The same conclusion can be made considering matrix form relations for $X(i)$
and $X(k)$. If both of them may satisfy the same two measurements then
\[
\left[
\begin{array}
[c]{c}%
y(0)\\
y(1)
\end{array}
\right]  =\left[
\begin{array}
[c]{cc}%
\psi_{i}(0) & \psi_{k}(0)\\
\psi_{i}(1) & \psi_{k}(1)
\end{array}
\right]  \left[
\begin{array}
[c]{c}%
X(i)\\
0
\end{array}
\right]
\]%
\begin{equation}
\left[
\begin{array}
[c]{c}%
y(0)\\
y(1)
\end{array}
\right]  =\left[
\begin{array}
[c]{cc}%
\psi_{i}(0) & \psi_{k}(0)\\
\psi_{i}(1) & \psi_{k}(1)
\end{array}
\right]  \left[
\begin{array}
[c]{c}%
0\\
X(k)
\end{array}
\right]  . \label{DET_ILL_CS}%
\end{equation}
Subtraction of the previous matrix equations results in
\[
\left[
\begin{array}
[c]{cc}%
\psi_{i}(0) & \psi_{k}(0)\\
\psi_{i}(1) & \psi_{k}(1)
\end{array}
\right]  \left[
\begin{array}
[c]{c}%
X(i)\\
-X(k)
\end{array}
\right]  =0.
\]
For $\psi_{i}(0)\psi_{k}(1)-\psi_{i}(1)\psi_{k}(0)\neq0$ follows
$X(i)=X(k)=0$. Therefore two different nonzero solutions $X(i)$ and $X(k)$ in
this case cannot exist. This concludes the proof that the solution is unique
if
\[
\psi_{i}(0)\psi_{k}(1)-\psi_{i}(1)\psi_{k}(0)=\det\left[
\begin{array}
[c]{cc}%
\psi_{i}(0) & \psi_{k}(0)\\
\psi_{i}(1) & \psi_{k}(1)
\end{array}
\right]  \neq0
\]
for any $i\neq k$. It also means that $\mathrm{rank}\left(  \mathbf{A}%
_{2}\right)  =2$ for any $\mathbf{A}_{2}$ being a $2\times2$ submatrix of the
matrix of coefficients (measurement matrix) $\mathbf{A}$. 

Let us consider $M$ measurements in this example. 
Since we have assumed that only one coefficient $X(i)$ is nonzero it will satisfy all measurements 
\[
\psi_{i}(0)X(i)=y(0)\text{, \ \ }\psi_{i}(1)X(i)=y(1), \ldots, ~\psi_{i}(M-1)X(i)=y(M-1).
\]
The solution will not be unique if there is another coefficient $X(k)$, $k\ne i$ satisfying
\[
\psi_{k}(0)X(k)=y(0)\text{, \ \ }\psi_{k}(1)X(k)=y(1), \ldots, \psi_{k}(M-1)X(k)=y(M-1).
\]
Then the corresponding coefficients of the measurement matrix satisfy
\[
\frac{\psi_{i}(0)}{\psi_{k}(0)}= \frac{\psi_{i}(1)}{\psi_{k}(1)} = \ldots = \frac{\psi_{i}(M-1)}{\psi_{k}(M-1)}.
\]
In this case measurement matrix is
\[
\mathbf{A=}\left[
\begin{array}
[c]{cccc}%
\psi_{0}(0) & \psi_{1}(0) & ... & \psi_{N-1}(0)\\
\psi_{0}(1) & \psi_{1}(1) & ... & \psi_{N-1}(1) \\
\vdots & \vdots & \ddots & \vdots \\
\psi_{0}(M-1) & \psi_{1}(M-1) & ... & \psi_{N-1}(M-1)
\end{array}
\right]
\]
The solution is not unique if any two columns are linearly dependent. The uniqueness requires that all two column submatrices $\mathbf{A}_2$ of $\mathbf{A}$ are of rank $2$. The determinant for all $\mathbf{A}_2^T\mathbf{A}_2$ is nonzero.

In numerical and practical applications we would not be satisfied, if for
example $\det(\mathbf{A}_2^T\mathbf{A}_2)\neq0$ but $\det(\mathbf{A}_2^T\mathbf{A}_2)=\varepsilon\ $close to zero. In this
case the theoretical condition for a unique solution would be satisfied,
however the analysis and possible inversion would be highly sensitive to any
kind of noise, including quantization noise. Thus, a practical requirement is
that the determinant is not just different from zero, but that it sufficiently
differs from zero so that an inversion stability and robustness to a noise is
achieved. Inversion stability for a matrix $\mathbf{B}=\mathbf{A}_2^T\mathbf{A}_2$ is commonly described
by the condition number of matrix
\[
\mathrm{cond}\left\{  \mathbf{B}\right\}  =\frac{\lambda_{\max}}{\lambda
_{\min}}%
\]
where $\lambda_{\max}$ and $\lambda_{\min}$ are the largest and the smallest
eigenvalue of matrix $\mathbf{B}$. 
%(when $\mathbf{B}^{H}\mathbf{B=BB}^{H}$)
The inversion stability worsens as
$\lambda_{\min}$ approaches to zero (when $\lambda_{\min}$ is small as
compared to $\lambda_{\max}$). For stable and robust calculations a
requirement
\[
\frac{\lambda_{\max}}{\lambda_{\min}}\leq1+\delta
\]
is imposed, with a nonnegative constant $\delta$ being sufficiently small. In
our example this condition should hold for all submatrices $\mathbf{A}_{2}$.

As a next example consider a signal described by a weighted sum of $K$
harmonics from a set of possible oscillatory functions $e^{j2\pi kn/N}$,
$k=0,$ $1,$ $2,$ $...,N-1$,%
\[
x(n)=A_{1}e^{j2\pi k_{1}n/N}+A_{2}e^{j2\pi k_{2}n/N}+...+A_{K}e^{j2\pi
k_{K}n/N},
\]
with $K\ll N$. In the DFT domain this signal will be sparse with
$X(k)=\mathrm{DFT}\left\{  x(n)\right\}  $ having only few nonzero values at
$k=k_{i}$, $i=1,2,...,K$. According to the sampling theorem the sampling of
this kind of signals should be adjusted to the maximal expected signal
frequency $k=\max\{k_{1},k_{2},...,k_{K}\}$. For an arbitrary set of
frequencies, it means that we should adjust sampling to the maximal
possible frequency $k=N-1$ and to use the full set of $N$ signal
values/measurements at $n=0,1,2,...,N-1$ in order to avoid aliasing. 

However, if
we know that the signal consists of only $K\ll N$ functions with unknown
amplitudes, then regardless of their frequencies, the signal can be fully
reconstructed from a reduced set of samples. Samples can be considered as
weighted measurements of the sparse function $X(k)$,%
\[
y(0)=x(n_{1})=\sum_{k=0}^{N-1}X(k)\psi_{k}(n_{1}),
\]
with the weighting coefficients $\psi_{k}(n_{1})=\exp(j2\pi n_{1}k/N)/N$. The
previous relation is the IDFT. Now a similar analysis like in the previous
illustrative example can be performed, assuming for example $K=1$ or $K=2$. We
can find the position and the value of nonzero $X(k)$ using just a few signal
samples $y(i)$. 

\begin{figure}[tb]
\begin{center}
\includegraphics{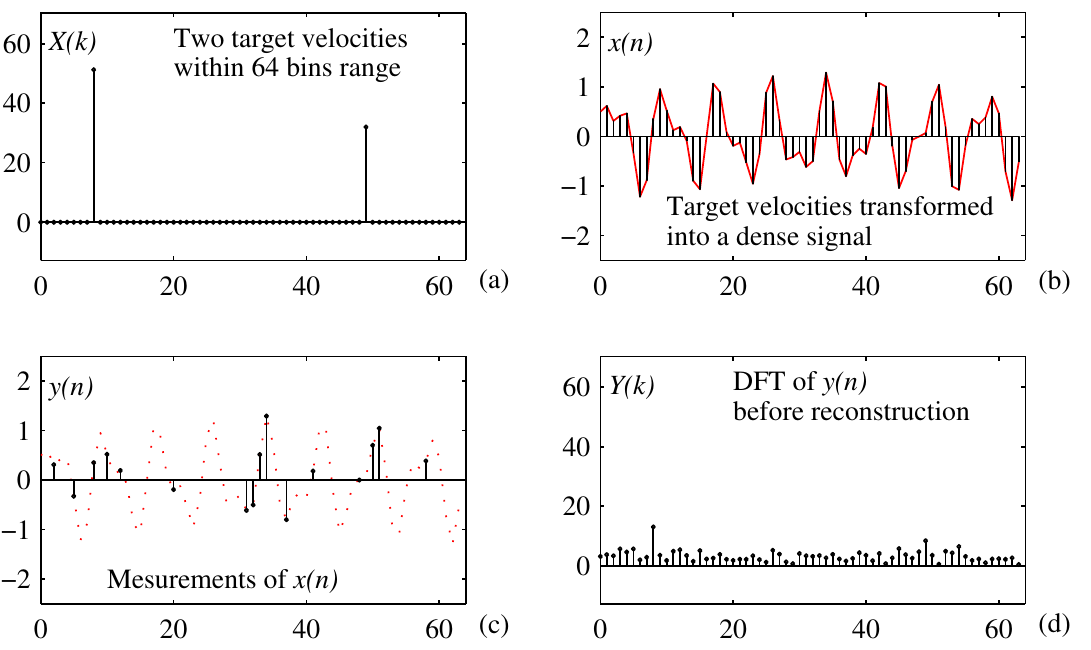}%
\caption{(a) Signal in the frequency domain, where it is sparse (for example, velocities of
two targets in Doppler radar signal). (b) Signal in the time domain, where it
is dense. (c) Reduced set of measurements (samples) and (d) its DFT before
reconstruction, calculated using the available samples only. Real parts of
signals are presented.}%
\label{radar_cs_ill}%
\end{center}
\end{figure}

This model corresponds to many signals in real life. For
example, in the Doppler-radar systems the speed of a radar target is
transformed into a frequency of a sinusoidal signal \cite{RadarCS,radarCS3}. Since the returned signal
contains only one or just a few targets, the signal representing target
velocity is a sparse signal in the DFT domain. It can be reconstructed from
fewer samples than the total number of radar return signal samples
$N$, Fig.\ref{radar_cs_ill}.%

%TCIMACRO{\FRAME{ftbpFU}{4.2133in}{2.5131in}{0pt}{\Qcb{(a) Signal in the
%frequency domain, where it is sparse (velocities of two targets in Doppler
%radar signal). (b) Signal in the time domain, where it is dense. (c) Reduced
%set of measurements (samples) and (d) its DFT before reconstruction,
%calculated using the available samples only. Real parts of signals are
%presented.}}{\Qlb{radar_cs_ill}}{radar_cs_ill.eps}%
%width 4.2133in;  height 2.5131in;  depth 0pt;  original-width 3.9842in;
%original-height 2.4591in;  cropleft "0";  croptop "1";  cropright "1";
%cropbottom "0";  filename 'Radar_CS_ILL.eps';file-properties "XNPEU";}}}%
%BeginExpansion
%EndExpansion

In signal processing the DFT as the domain of signal sparsity is commonly used,
since it plays the central role in engineering applications \cite{Wiley,Uniq}. Note that in the
compressive sensing theory random measurement matrices are mainly used.  The compressive
sensing results and algorithms are used as a tool to solve the
problems involving sparse signals.

\section{Definitions}

A big set of discrete-time data $x(n), n=0,1,\dots,N-1,$ with a large number
of samples $N$ is considered. Its coefficients in a transformation domain
are denoted as $$\mathbf{X}=[X(0), ~X(1),...,~X(N-1)]^T,$$ where $T$ represents
the transpose operation. We consider a signal to be sparse in this transformation
domain if the number of nonzero coefficients $K$ is much smaller than the
number of the original signal samples $N$, i.e., if
$X(k)=0$
for 
$
k\notin \mathbb{K}=\{k_{1},k_{2},...,k_{K}\}$ and $K\ll N$. The number of
nonzero coefficients is commonly denoted by $\left\Vert \mathbf{X}\right\Vert
_{0} $
\[
\left\Vert \mathbf{X}\right\Vert _{0}=\mathrm{card}\left\{  \mathbb{K}
\right\}  =K,
\]
where $\mathrm{card}\left\{  \mathbb{K}\right\}$ is the cardinality of set
$\mathbb{K}.$ It is equal to the
number of elements in $\mathbb{K}$. It is called  the $\ell_{0}$-norm (norm-zero)
or  the $\ell_{0}$-pseudo-norm of vector $\mathbf{X}$ although it does not
satisfy the norm properties. 

The observations/measurements are defined as linear combinations of signal
coefficients in the sparsity domain
\begin{equation}
y(m)=\sum_{k=0}^{N-1}X(k)\psi_{k}(m), \label{Hip00}%
\end{equation}
where $m=0,1,\ldots, M-1$ is the measurement index and $\psi_{k}(m)$ are
the weighting coefficients. 
The vector form of the measurement signal is denoted by $\mathbf{y}$
\[
\mathbf{y=[}y(0),~y(1),~...,y(M-1)]^{T}.
\]

The measurements defined by Eq.(\ref{Hip00}) can be written as a undetermined system of
$M<N$ equations
\[
\begin{bmatrix}
y(0)\\
y(1)\\
\vdots\\
y(M-1)
\end{bmatrix}
  =
\begin{bmatrix}
\psi_{0}(0) & \psi_{1}(0) & \cdots & \psi_{N-1}(0)\\
\psi_{0}(1) & \psi_{1}(1) & \cdots & \psi_{N-1}(1)\\
\vdots & \vdots &  \ddots & \vdots\\
\psi_{0}(M-1) & \psi_{1}(M-1) & \cdots & \psi_{N-1}(M-1)
\end{bmatrix}
\begin{bmatrix}
X(0)\\
X(1)\\
\vdots\\
X(N-1)
\end{bmatrix}
\]
or using matrix notation 
\[
\mathbf{y=AX}
\]
where $\mathbf{A}$ is the measurement matrix of size $M\times N$.
\index{Measurement Matrix}

The fact that the signal is sparse with $X(k)=0$ for $k\notin \mathbb{K}=\{k_1,k_2,...,k_K\}$
is not included in the measurement matrix
$\mathbf{A}$ since the positions of the nonzero values are unknown. If the
knowledge that $X(k)=0$ for $k\notin \mathbb{K}$ were
included then a reduced system would be obtained as
\[
\begin{bmatrix}
y(0)\\
y(1)\\
\vdots\\
y(M-1)
\end{bmatrix}
=
\begin{bmatrix}
\psi_{k_{1}}(0) & \psi_{k_{2}}(0) & \cdots & \psi_{k_{K}}(0)\\
\psi_{k_{1}}(1) & \psi_{k_{2}}(1) & \cdots & \psi_{k_{K}}(1)\\
\vdots & \vdots & \ddots & \vdots\\
\psi_{k_{1}}(M-1) & \psi_{k_{2}}(M-1) & \cdots  & \psi_{k_{K}}(M-1)
\end{bmatrix}
\begin{bmatrix}
X(k_{1})\\
X(k_{2})\\
\vdots\\
X(k_{K})
\end{bmatrix}
\]
with a reduced $M\times K$ measurement matrix $\mathbf{A}_{K}$ defined as
\begin{equation}
\mathbf{y}=\mathbf{A}_{K}\mathbf{X}_{K}.\label{A_K}
\end{equation}
This is an overdetermined system of equation, $K<N$. Matrix $\mathbf{A}_{K}$ would be formed if we assumed/knew the positions
of nonzero
samples $k\in \mathbb{K}$. It would follow from the measurement matrix $\mathbf{A}$
by omitting the columns corresponding to the zero-valued coefficients in
$\mathbf{X}$.

\subsection{Common Measurement Matrices}

Some common measurement matrices used in practical applications and theoretical
considerations will be presented here.

Randomness of measurement matrices is a favorable property in compressive
sensing and matrices with random elements are often used. The most common
is the measurement matrix with zero-mean unity variance Gaussian distributed
numbers as elements
$$\phi_k(n) \sim \frac{1}{\sqrt{M}} \mathcal{N}(0,1) $$ 
normalized with $1/\sqrt M$ so that the energy of each column is one.

In signal processing the most common transform is the DFT. The coefficients
of its direct transform matrix 
$\mathbf{\Phi}$ are defined as
$$\phi_k(n) =\exp(-j2\pi nk/N).$$
 The inverse DFT matrix coefficients are $\psi_k(n)
=\frac{1}{N}\exp(j2\pi nk/N)$. Commonly the measurements are the signal samples
$y(m-1)=x(n_m)$ for $m=1,\ldots,M$ where $$n_m \in \mathbb{M}=\{n_1,n_2,\dots,n_{M}\}\subset\{0,1,\dots,N-1\},$$
and
$$ y(m-1)=x(n_m)=\frac{1}{N}\sum_{k=0}^{N-1}X(k)e^{j2\pi n_mk/N}.$$
Therefore, the measurement matrix is obtained by keeping the rows of the
inverse DFT matrix corresponding to the samples at $n_m\in \{0,1,\dots,N-1\}$,
for the measurements $m=1,2,\ldots,M$,
\begin{equation}
\mathbf{A}=\frac{1}{N}
\begin{bmatrix}
1 & e^{j2\pi n_1/N} & \cdots & e^{j2\pi n_1(N-1)/N}\\
1 & e^{j2\pi n_2/N} & \cdots & e^{j2\pi n_2(N-1)/N}\\
\vdots & \vdots & \ddots & \vdots\\
1 & e^{j2\pi n_{M}/N} & \cdots  & e^{j2\pi n_{M}(N-1)/N}
\end{bmatrix}.\label{FTp}
\end{equation}
This is a partial inverse DFT matrix.
In compressive sensing theory it is common to normalize the measurement matrix
so that the energy of its columns (diagonal elements of $\mathbf{A^H}\mathbf{A}$
matrix) is equal to one. Then the factor $1/N$ in $\mathbf{A}$ should be
replaced by $1/\sqrt M$.

In order to increase randomness in the Fourier transform matrix, the measurements
may be taken at any random instant. Then the measurement vector elements
are $y(m-1)=x(t_m)$ where $t_m$, $m=1,2,\dots,M$ are random instants within
the considered time interval $T$. The measurement matrix follows then from
the Fourier series definition $x(t)=\sum_{k=0}^{N-1}X(k)\exp(j2\pi k t/T)$.
It has been assumed that the Fourier series coefficients are within $0 \le
k \le N-1$. The measurements matrix is     
\begin{equation}
\mathbf{A}=
\begin{bmatrix}
1 & e^{j2\pi t_1/T} & \cdots & e^{j2\pi t_1(N-1)/T}\\
1 & e^{j2\pi t_2/T} & \cdots & e^{j2\pi t_2(N-1)/T}\\
\vdots & \vdots & \ddots & \vdots\\
1 & e^{j2\pi t_{M}/T} & \cdots & e^{j2\pi t_{M}(N-1)/T}
\end{bmatrix} \label{pFT}
\end{equation}
with a possible normalization factor $1/\sqrt M$.
This measurement matrix is a partial random inverse Fourier transform matrix.

\section{Reconstruction Problem Formulation}

The signal can be reconstructed from its measurements defined by vector
$\mathbf{y}$ by finding the sparsest vector $\mathbf{X}$ that corresponds
to the measurements $\mathbf{y}$. Hence, by introducing the notation for the number
of components based on the $\ell _0$-norm  $K=\left\Vert \mathbf{X}\right\Vert
_{0}$,  the fundamental minimization problem can be formulated as: 
\begin{equation}
\min\left\Vert \mathbf{X}\right\Vert _{0}  \text{ subject to  } \mathbf{y}=\mathbf{A}\mathbf{X}.
\end{equation}

In general, the $\ell _0$-norm is not very suitable for most minimization
methods. However, a 
class of algorithms is based on the minimization of 
the number of coefficients $K=\left\Vert \mathbf{X}\right\Vert _{0}$ in an
implicit way. For instance, in certain applications we can predict the number
of components or we are able to estimate the position of non-zero coefficients.
Thus, compared to the direct search method, the computational complexity
will be significantly reduced if we
are able to estimate positions of nonzero coefficients and solve the problem
with the minimal possible number of 
nonzero coefficients. The important class of this algorithms are \textit{matching pursuit} (MP) algorithms. 

Minimization of the number of non-zero coefficients using
the $\ell_{0}$-norm is a nonconvex optmization problem that cannot be solved
using well developed iterative algorithms and linear programming methods \cite{IntroR,MP,OMP_2,CoSaMP_1,LLASSO,MS_1,mallat1993,FISTA1}.
To avoid dealing with NP-hard problems, significant efforts have
been undertaken to replace the nonconvex and discontinuous $\ell_{0}$-norm
with a convex and continuous norm that would be more appropriate for optimization.
As a result, the $\ell_{1}$-norm (norm-one) has been commonly employed in
many signal reconstruction approaches \cite{GradBT,Gradient1,GradientIET,L1Magic}. It has been shown that, under certain
conditions, minimization of the $\ell_{1}$-norm produces the same solution
as the minimization of  the $\ell_{0}$-norm.

In the $\ell_{1}$-norm based reconstructions the problem is formulated as%

\[
\min\left\Vert \mathbf{X}\right\Vert _{1}\text{ \ \ \ subject to
\ \ }\mathbf{y=AX}%
\]
where $$\left\Vert \mathbf{X}\right\Vert _{1}=\sum_{k=0}^{N-1}\left\vert
X(k)\right\vert. $$ 
This is the so-called \textit{basis pursuit} (BP) approach to sparse signal
reconstruction. 

\subsection{Conditions for Unique Reconstruction} 

\subsubsection{Spark}

The spark of a matrix $\mathbf{A}$ is defined as the smallest number of linearly dependent columns of $\mathbf{A}$. In other words if $\operatorname{spark}\{\mathbf{A}\}=K$, then any collection of $K_1<K$ columns of $\mathbf{A}$ are linearly independent.

Spark can also be defined as a minimal number of nonzero entries in a vector $\mathbf{X}\ne \mathbf{0}$ such that $\mathbf{A}\mathbf{X}=\mathbf{0}$
$$
\min \Vert \mathbf{X} \Vert_0 \qquad \text{such that } \mathbf{A}\mathbf{X}=\mathbf{0}
$$

If matrix $\mathbf{A}$ is of size $N \times K$ with $N>K$ and $\operatorname{spark}\{\mathbf{A}\}=K+1$ then all $K \times K$ submatrices  of matrix $\mathbf{A}$ are nonsingular, i.e., with nonzero determinant. 

The analysis of a signal with an arbitrary sparsity $K$ is similar to the
analysis for $K=1$. To get the first set of possible
solutions for $K$ nonzero coefficients (of sparsity $K$) we need $K$ measurements.
For any combination of $K$ (out of $N$) nonzero coefficients $X(k)$, $k\in\{k_{1},k_{2},...,k_{K}\}$,
we will get a possible solution. There exist $\binom{N}{K}$ such possible
combinations/solutions. Additional $K$ measurements
will be used to produce another set of  $\binom{N}{K}$ possible solutions.
The intersection of these two sets is then the solution of our problem.

\textit{Consider the case when the number of measurements $M$ is twice higher than the sparsity $K$, $M=2K$. The $K$-sparse solution is unique if the determinants of all $\mathbf{A}_{2K}$ submatrices
of matrix $\mathbf{A}$ are different from zero.} 

This statement will be proven
by contradiction.  Assume that $M=2K$ measurements are available  within
the vector $\mathbf{y}$. Assume that two different solutions for $\mathbf{X}$
 of sparsity $K$ exist.  Denote the nonzero parts of the solutions by $\mathbf{X}_K^{(1)}$
and $\mathbf{X}_K^{(2)}$. Both of them satisfy the measurements equation,

$$\mathbf{A}_K^{(1)}\mathbf{X}_K^{(1)}=\mathbf{y}$$ and 
$$\mathbf{A}_K^{(2)}\mathbf{X}_K^{(2)}=\mathbf{y},$$ where $\mathbf{A}_K^{(1)}$
and $\mathbf{A}_K^{(2)}$ are two different submatrices of matrix $\mathbf{A}$
of size $M\times K$ corresponding to the elements in $\mathbf{X}_K^{(1)}$
and $\mathbf{X}_K^{(2)}$. If we rewrite these equations by adding zeros
\begin{gather}
\begin{bmatrix}\mathbf{A}_K^{(1)} ~~~ \mathbf{A}_K^{(2)}
\end{bmatrix}\begin{bmatrix}\mathbf{X}_K^{(1)} \\ \mathbf{0}_K \end{bmatrix}
=\mathbf{y} \text{~~~~~~~and~~~~~~~}
\begin{bmatrix}\mathbf{A}_K^{(1)} ~~~ \mathbf{A}_K^{(2)}
\end{bmatrix}\begin{bmatrix} \mathbf{0}_K \\\mathbf{X}_K^{(2)}\end{bmatrix}
=\mathbf{y}
\end{gather} 
and subtract them we get
\begin{gather}
\begin{bmatrix}\mathbf{A}_K^{(1)} ~~~ \mathbf{A}_K^{(2)}
\end{bmatrix}\begin{bmatrix}\mathbf{X}_K^{(1)} \\ -\mathbf{X}_K^{(2)} \end{bmatrix}
=\mathbf{0}. \label{SYS2K}
\end{gather} 
There are no nonzero solutions for $\mathbf{X}_K^{(1)}$ and $\mathbf{X}_K^{(2)}$
if the determinant of matrix $\mathbf{A}_{2K}= \begin{bmatrix}\mathbf{A}_K^{(1)}
~~~ \mathbf{A}_K^{(2)}
\end{bmatrix}$ is nonzero. If all possible submatrices $\mathbf{A}_{2K}$
(including all lower order submatrices) of measurement matrix $\mathbf{A}$
are nonsingular then two solutions of sparsity $K$ cannot exist, and the
solution is unique.  Note that there are $\binom{N}{2K}$ submatrices $\mathbf{A}_{2K}$.

Based on the previous analysis, the solution for a $K$ sparse problem is
unique if 
$$\operatorname{spark}\{\mathbf{A}\}>2K.$$

For $M>2K$ the matrix $\mathbf{A}_{2K}= \begin{bmatrix}\mathbf{A}_K^{(1)}
~~~ \mathbf{A}_K^{(2)}
\end{bmatrix}$ dimension is $M \times 2K$. Again if $\mathrm{rank} (\mathbf{A}_{2K})=\mathrm{rank} (\mathbf{A}^T_{2K}\mathbf{A}_{2K})=2K$  system  (\ref{SYS2K}) does not have a nonzero solution. It means that the reconstruction is unique. If $\mathrm{rank} (\mathbf{A}_{2K})=\mathrm{rank}
(\mathbf{A}^T_{2K}\mathbf{A}_{2K})=2K$ for all submatrices $\mathbf{A}_{2K}$ then $\operatorname{spark}\{\mathbf{A}\}>2K.$   

\textit{If the vector $\mathbf{X}$ is of sparsity $K$, with
$
\left\Vert \mathbf{X}\right\Vert _{0}=K
$
then if
\[
K<\frac{1}{2}\mathrm{spark}\left\{  \mathbf{A}\right\}
\]
the solution $\mathbf{X}$ is unique.
}

In order to prove this statement (that has been already explained) consider a measurement matrix $\mathbf{A}$ whose spark is
$\mathrm{spark}\left\{  \mathbf{A}\right\}  $. Then for a sparse vector
$\mathbf{X}$ of sparsity $K=$ $\mathrm{spark}\left\{  \mathbf{A}\right\}  $
obviously there exists such a combination of nonzero elements in $\mathbf{X}$
so that they coincide with the dependent columns. Then we can obtain%
\[
\mathbf{AX=0.}%
\]
This property is used for the spark definition as well.

Note that for any $\mathbf{X}$ of sparsity $K<\mathrm{spark}\left\{
\mathbf{A}\right\}  $ the relation $\mathbf{AX=0}$ will not hold, since
corresponding independent columns of $\mathbf{A}$ multiplied with nonzero  elements of $\mathbf{X}$ cannot produce a zero result. Since $K<\mathrm{spark}\left\{
\mathbf{A}\right\}  $ it means that all sets of $K$ columns from $\mathbf{A}$ are independent.

The proof of the previous statement will be 
based on the contradiction.
Assume that $\mathbf{X}$ is a solution of $\mathbf{AX=y}$ and that its sparsity satisfies $K<\frac{1}%
{2}\mathrm{spark}\left\{  \mathbf{A}\right\}  $. Assume also that there is another
solution $\mathbf{H}$ such that $\mathbf{AH=y}$ and $\mathbf{H}$ is also sparse with
sparsity lower than $\frac
{1}{2}\mathrm{spark}\left\{  \mathbf{A}\right\}  $.
Since
\begin{gather*}
\mathbf{AH=AX=y}\\
\mathbf{A(H-X)=0}%
\end{gather*}
then
\[
\mathrm{spark}\left\{  \mathbf{A}\right\}  < \left\Vert \mathbf{H}%
-\mathbf{X}\right\Vert _{0}
\]
or
\begin{gather*}
\mathrm{spark}\left\{  \mathbf{A}\right\}  <\left\Vert \mathbf{H-X}\right\Vert
_{0}\leq\left\Vert \mathbf{H}\right\Vert _{0}+\left\Vert \mathbf{X}\right\Vert_{0} \\
\mathrm{spark}\left\{  \mathbf{A}\right\}  -\left\Vert \mathbf{H}\right\Vert
_{0}\leq\left\Vert \mathbf{X}\right\Vert _{0}.
\end{gather*}
The inequality follows from the fact that two nonzero elements, at the same position in $\mathbf{H}$ and $\mathbf{X}$, can produce a zero element in $\mathbf{H-X}$, while two zero elements in these vectors cannot produce a nonzero element  in $\mathbf{H-X}$. If there is another solution $\mathbf{H}$ such that $\left\Vert \mathbf{H}%
\right\Vert _{0}<\frac{1}{2}\mathrm{spark}\left\{  \mathbf{A}\right\}  $ then
from the last inequality follows that $\left\Vert \mathbf{X}\right\Vert _{0}%
> \frac{1}{2}\mathrm{spark}\left\{  \mathbf{A}\right\}  .$ This is a
contradiction to the assumption that both solutions $\mathbf{H}$ and
$\mathbf{X}$ have sparsity lower than $\frac{1}{2}\mathrm{spark}\left\{
\mathbf{A}\right\}  $.

\subsubsection{Restricted Isometry Property}

Note that for any square matrix its determinant is equal to the product of
its eigenvalues
$$\det\{\mathbf{A}^T_{2K}\mathbf{A}_{2K}\}=\lambda_1 \lambda_2 \cdot \ldots \cdot \lambda_{2K}.
$$
The condition that the solution is unique if the determinant of $\mathbf{A}^T_{2K}\mathbf{A}_{2K}$ for all $\mathbf{A}_{2K}$
submatrices
of matrix $\mathbf{A}$ are different from zero can be rewritten as
$$ \min_{i}|\lambda_i|>0.$$

In numerical and practical applications we would not be satisfied, if any
of the determinants is very close to zero. In this
case the theoretical condition for a unique solution would be satisfied,
however the analysis and possible inversion would be highly sensitive to
any
kind of noise in measurements. Thus, a practical requirement
is
that the determinant is not just different from zero, but that it sufficiently
differs from zero so that an inversion stability and noise robustness
is
achieved.

From the matrix theory it is known that the norm of a matrix  $\mathbf{A}_{2K}$
satisfies
\begin{equation}
\lambda_{\min} \le 
\frac{\left\| \mathbf{A}_{2K}\mathbf{X}_{2K}\right\|_2^2}%
{\left\| \mathbf{X}_{2K} \right\|_2^2}
=\frac{\mathbf{X}_{2K}^T\mathbf{A}_{2K}^T\mathbf{A}_{2K}\mathbf{X}_{2K}}
{\mathbf{X}_{2K}^T\mathbf{X}_{2K}}\le \lambda_{\max},
\label{normA} \end{equation} 
where $ \lambda_{\min}$ and $\lambda_{\max}$ are the minimal and the maximal
eigenvalue of the matrix $
\mathbf{A}_{2K}^T
\mathbf{A}_{2K}
$  and ${\left\| \mathbf{X}_{}
\right\|_2^2}=|X(0)|^2+\dots|X(N-1)|^2$ is the squared $\ell_{2}$-norm (norm-two).

The isometry property for a linear transformation matrix $\mathbf{A}$ holds
if 
$$\left\| \mathbf{A}\mathbf{X}\right\|_2^{2}=\left\| \mathbf{X} \right\|_2^{2}
~~~\text{or}~~~~\frac{\left\| \mathbf{A}\mathbf{X}\right\|_2^{2}}
{\left\| \mathbf{X} \right\|_2^{2}}=1.$$ 

The restricted isometry property
(RIP) for a matrix $\mathbf{A}_{2K}$ and a $2K$-sparse vector $\mathbf{X}_{2K}$
holds if
\begin{equation}
1-\delta_{2K} \le 
\frac{\left\| \mathbf{A}_{2K}\mathbf{X}_{2K}\right\|_2^{2}}%
{\left\| \mathbf{X}_{2K} \right\|_2^{2}}
\le 1+\delta_{2K},
\label{eq:RIP}
\end{equation} 
where $ 0 \le \delta_{2K}<1 $ is the isometric constant.
From Eqs.(\ref{normA}) and (\ref{eq:RIP}) we can write
$$
\delta_{2K}= \max\{ 1-\lambda_{\min},\ \lambda_{\max}-1\}.
$$
Commonly, isometric constant is defined by $\lambda_{\max}-1$ and it is calculated
as maximal eigenvalue of matrix
$
\mathbf{A}_{2K}^T
\mathbf{A}_{2K}
-\mathbf{I}$. Normalized energies of the columns of matrix $
\mathbf{A}$ (diagonal elements of $
\mathbf{A}_{2K}^T
\mathbf{A}_{2K}
$) are assumed. Otherwise, the normalization factors should be added. For
complex-valued matrices Hermitian transpose should be used in $
\mathbf{A}_{2K}^H
\mathbf{A}_{2K}
$.  

\textit{For a $K$-sparse vector $\mathbf{X}$ and a measurement matrix $\mathbf{A}$
the RIP is satisfied if relation (\ref{eq:RIP}) holds for all submatrices
$\mathbf{A}_{K}$ with $ 0 \le \delta_{K}<1 $.
The solution for $K$-sparse vector is unique if the measurement matrix satisfy
the RIP for $2K$-sparse vector $\mathbf{X}$ with  $ 0 \le \delta_{2K}<1 $.
}

Note that if the RIP is satisfied then $\lambda_{\min}>0$. It means that there is no $\mathbf{A}_{2K}$ submatrix of $\mathbf{A}$ such that  $
\mathbf{A}_{2K}^H
\mathbf{A}_{2K}
$ is a singular matrix. The uniqueness proof reduces to the previous one.

Restricted isometry property for small $\delta_{2K}$ is closer to the isometry
property and improves the solution stability. It can be related   to the
 matrix conditional number. The conditional number of a matrix 
$\mathbf{A}^T_{2K}\mathbf{A}_{2K}$ is defined as the ratio of  its maximal
and minimal eigenvalues

\[
\mathrm{cond}\left\{  \mathbf{A}^T_{2K}\mathbf{A}_{2K} \right\}  =\frac{\lambda_{\max}}{\lambda_{\min}}.
\]
If a matrix $\mathbf{A}_{2K}$ satisfies the restricted isometry property
with $\delta_{2K}$ then
$$
\mathrm{cond}\left\{\mathbf{A}^T_{2K}  \mathbf{A}_{2K}\right\} \le \frac{1+\delta_{2K}}{1-\delta_{2K}}.
$$
With small values of $\delta_{2K}$ the conditional number is close to one,
meaning stable invertibility and low sensitivity to the input noise (small
variations of the input signal (measurements) do not cause large variations
of the result).
Common requirement for this constant is
$$0 \le \delta_{2K}<\sqrt2-1.$$
The restricted isometry constant within this range will also guarantee the equivalence of the solutions obtained in the reconstruction based on the  $\ell_{0}$-norm and the $\ell_{1}$-norm minimization \cite{RIP,RIPP}.

\subsubsection{Coherence}

The mutual coherence (coherence
index) of a matrix $\mathbf{A}$ is defined as the maximal absolute value
of
the normalized scalar product of its two columns%
\index{Coherence}
\[
\mu=\max\left\vert \mu_{mk}\right\vert \text{, for }m\neq k
\]
where
\begin{equation}
\mu_{mk}=\frac{\sum_{i=0}^{M-1}\alpha_{m}(i)\alpha_{k}^{\ast}(i)}
{\sqrt{\sum_{i=0}^{M-1}\left\vert
        \alpha_{m}(i)\right\vert ^{2}\sum_{i=0}^{M-1}\left\vert
        \alpha_{k}(i)\right\vert ^{2}}}
\end{equation}
and $\alpha_{k}(i)$ are the elements of the $k$th column of matrix $\mathbf{A}$. If $\sum_{i=0}^{M-1}\left\vert
        \alpha_{k}(i)\right\vert ^{2}=\sum_{i=0}^{M-1}\left\vert
        \alpha_{m}(i)\right\vert ^{2}$ then 
\begin{equation}
\mu_{mk}=\frac{\sum_{i=0}^{M-1}\alpha_{m}(i)\alpha_{k}^{\ast}(i)}
{{\sum_{i=0}^{M-1}\left\vert
        \alpha_{k}(i)\right\vert ^{2}}}.
\end{equation}        

Note that $\mu_{mk}$, $m \ne k$, are the off-diagonal elements of matrix $\mathbf{A}^H\mathbf{A}$ normalized with the corresponding diagonal elements.

This index plays
an important role in the analysis of measurement matrices. The coherence
index should be as small as possible, or in other words the incoherence is
a desirable property. With smaller values of coherence index
the matrix $\mathbf{A}^H\mathbf{A}$ is closer to the identity matrix. 

The condition that all eigenvalues of matrix $\mathbf{A}^H\mathbf{A}$ are nonzero can be written in terms of the coherence index. In general, the eigenvalue relation for matrix $\mathbf{A}^H\mathbf{A}$ reads   
$$(\mathbf{A}^H\mathbf{A})\mathbf{u}=\lambda\mathbf{u}$$
where $\mathbf{u}$ denotes an eigenvector. For an eigenvector we can always choose that its maximal coordinate is $u_m=\max_k(u_k)=1$ and $u_k \le 1$ for $k \ne m$. Now we can write the general eigenvalue relation in the form
$$\sum_k\mu_{mk}u_k=\lambda u_m =\lambda$$
or   
$$\sum_{k,k\ne m}\mu_{mk}u_k=\lambda -\mu_{mm}$$
From this relation we can conclude 
$$|\lambda -\mu_{mm}| \le \sum_{k,k\ne m}|\mu_{mk}u_k|\le \sum_{k,k\ne m}|\mu_{mk}|.$$
Considering eigenvalue $\lambda$ as a variable and $\mu_{mk}$ as constants we conclude that the last inequality describes a disc area with the center at $\mu_{mm}$ and radius $ \sum_{k,k\ne m}|\mu_{mk}|$. It does not include point $\lambda=0$ if
\begin{equation}
 \mu_{mm}>\sum_{k,k\ne m}|\mu_{mk}|.\label{MIII}
 \end{equation}
Therefore the matrix $\mathbf{A}^H\mathbf{A}$ will be nonsingular if the above condition is met. This is the Gershgorin circle (disk) theorem. 

For normalized matrix $\mathbf{A}^H\mathbf{A}$ we have $\mu_{mm}=1$ and 
$$\mu=\max_{m \ne k}\left\vert \mu_{mk}\right\vert.$$ 

We have already concluded that the solution for $K$-sparse vector will be unique if for all possible submatrices $\mathbf{A}_{2K}$ the matrices $\mathbf{A}_{2K}^H\mathbf{A}_{2K}$ are nonsingular. Note that the off-diagonal elements of $\mathbf{A}_{2K}^H\mathbf{A}_{2K}$ are a subset of the off-diagonal elements of matrix $\mathbf{A}^H\mathbf{A}$. The same holds for the diagonal elements. It mean that the coherence $\mu$ of matrix $\mathbf{A}$ will be always greater than or equal to the coherence of any submatrix $\mathbf{A}_{2K}$.

\textit{The nonsingularity condition for all matrices $\mathbf{A}_{2K}^H\mathbf{A}_{2K}$, and unique solution for a $K$ sparse vector $\mathbf{X}$, is achieved if 
$$ 1>(2K-1)\mu$$
or 
$$ K<\frac{1}{2}(1+\frac{1}{\mu}).$$
}
The proof follows from (\ref{MIII})  
for normalized matrix $\mathbf{A}^H\mathbf{A}$. The inequality  
$$1=\mu_{mm}>\sum_{k=1,k\ne m}^{2K}|\mu_{mk}|$$    
is satisfied if $1>(2K-1)\mu$ since $\sum_{k=1,k\ne m}^{2K}|\mu_{mk}|<(2K-1)\mu.$ 

The coherence index cannot be arbitrarily small for an $M\times
N$
matrix
$\mathbf{A}$ ($M<N$). The Welch upper bound%
\index{Welch bound} relation holds%
\begin{equation}
\mu\geq\sqrt{\frac{N-M}{M\left(  N-1\right)  }}. \label{WelchBound}%
\end{equation}
The  Gershgorin circle (disk) theorem can be used to determine the spark lower bound. If the relation $K<(1+1/\mu)/2$ holds for a given $K$ then it holds for any order lower than $K$. It means that all submatrices $\mathbf{A}_{2K}^H\mathbf{A}_{2K}$ of $2K$ and lower order are nonsingular. Therefore the spark of such a matrix is greater than $2K$  
 \begin{equation} 
\operatorname{spark}(\mathbf{A})>2K
\label{eq:coh2} 
\end{equation}
or 
$$K<\frac{1}{2}\operatorname{spark}(\mathbf{A})$$
if $K<(1+1/\mu)/2$.  It means that 
$$ \operatorname{spark}(\mathbf{A})\ge1+\frac{1}{\mu}.$$ 

\subsection{Numerical Example}
Consider $5\times 8$ measurement matrix
\begin{equation}
\mathbf{A}=
\left[
\begin{array}{rrrrrrrr}
  0.1 &   0.1 &   0.3 &  -0.7 &   0.7 &  -0.1 &   0.1 &   0.3 \\
  0.4 &  -0.8 &  -0.4 &  -0.1 &   0.3 &   0.3 &   0.3 &  -0.5 \\
  0.3 &   0.5 &  -0.5 &   0.4 &   0.5 &  -0.7 &   0.1 &  -0.4 \\
 -0.7 &  -0.3 &   0.1 &   0.3 &   0.4 &  -0.5 &   0.5 &  -0.7 \\
 -0.5 &   0.1 &  -0.7 &  -0.5 &  -0.1 &  -0.4 &  -0.8 &  -0.1 \\
\end{array} \right].
\end{equation}
Columns of this matrix are normalized. The norm of each column is equal to one. The matrix dimensions are small so we can solve NP-hard problems and calculate the spark and the restricted isometry property constants, by checking all possible combinations, in a reasonable time.

\subsubsection{Spark Calculation}
For the spark calculation we first concluded that there is no all zero column, meaning that $\operatorname{spark}\{\mathbf{A}\} >1 $. Then we have check linear dependence of each pairs of columns.   In total $8\choose{2}$ combinations are checked by calculating rank of each $2 \times 5$ submatrix. In all cases rank was equal to $2$ meaning that all pair of two columns are linearly independent. It means that  $\operatorname{spark}\{\mathbf{A}\} >2 $. Next all $8\choose{3}$ possible combinations of three columns are considered. For all submatrices we concluded that their rank is $3$ meaning that there is no a set of three linearly dependent columns in the measurement matrix $\mathbf{A}$. Therefore $\operatorname{spark}\{\mathbf{A}\} >3 $.  Calculation is repeated for all combinations of four and five columns with the same result. The final conclusion is that the spark of this matrix is $\operatorname{spark}\{\mathbf{A}\}=6$ meaning that all combinations of five and less columns are linearly independent. The uniqueness condition based on matrix spark state that sparsity $K$ limit is
$$ K<\frac{1}{2} \operatorname{spark}\{\mathbf{A}\}=3. $$
According to the spark based uniqueness condition, the reconstruction is unique for $K=1$ and $K=2$. We may conclude that if we find a sparse vector $\mathbf{X}$ in the reconstruction with sparsity  $K\le2$ then this is the sparsiest possible solution of our problem.

\subsubsection{Coherence Calculation}

Coherence of the considered matrix is calculated a maximal absolute value of the off-diagonal element of $\mathbf{A}^T\mathbf{A}$. If the diagonal elements were not normalized then this maximal value should be normalized with the diagonal values of this matrix. For the considered measurement matrix $\mathbf{A}$ we get

 $$\mu=0.49$$
  resulting in the sparsity limit
$$ K<\frac{1}{2}(1+\frac{1}{\mu}) \approx 1.5204. $$
The worst case, that determine value of $\mu$, was coherence between 5th and 7th column of the considered matrix. The reconstruction is unique only for $K=1$. Note that in contrast to the spark limit this condition will guarantee that the same unique solution is obtained using $\ell_1$-norm and $\ell_0$-norm. This is the reason why this limit is more conservative.

For a measurement matrix of order $5 \times 8$ the smallest possible value of the the coherence index is
$$\mu\geq\sqrt{\frac{N-M}{M\left(  N-1\right)  }}=\sqrt{\frac{3}{5 \times 7  }}=0.2928$$
with maximal possible bound $K<2.2078$. The matrix with minimal coherence index is quite specific and it is called the equiangular tight frame (ETF). In practice many optimization approaches are based on finding the measurement matrix with coherence as low as possible (as close to the ETF as possible).

\subsubsection{Restricted Isometry Property Constant Calculation}

Restricted isometry property (RIP) constants of orders $1$, $2$, $3$, $4$, and $5$ are calculated. For the calculation of the  RIP constant with assumed sparsity $K=1$ all possible submatrices $\mathbf{A}_1$ are formed. These are $5 \times 1$ matrices. There are $8$ of them. The matrices $\mathbf{A}_1^T\mathbf{A}_1$ are formed.    All of them are scalars equal to $1$ with $\lambda=1$, resulting in 
$$
\delta_{1}= \max\{ 1-\lambda_{\min},\ \lambda_{\max}-1\}=0.
$$
 Next the sparsity $K=2$ of the resulting $\mathbf{X}$ is assumed. All possible measurement submatrices $\mathbf{A}_2$ corresponding to this sparsity are formed. There are $8\choose 2$ of them. The matrices $\mathbf{A}_2^T\mathbf{A}_2$
are formed. Then their eigenvalues are calculated.  The RIP constant $\delta_2$ is obtained as a maximal value of $$
\delta_{2}= \max\{ 1-\lambda_{\min},\ \lambda_{\max}-1\}=0.49.
$$
over all possible submatrices $\mathbf{A}_2$. 

The calculation is repeated for assumed sparsity $K=3,4,5$ by forming corresponding submatrices $\mathbf{A}_3$, $\mathbf{A}_4$, and $\mathbf{A}_5$, respectively. The obtained numerical values for these sparsities are
\begin{align*}
\delta_3 & =0.9406 \\
\delta_4 & =1.2063 \\
\delta_5 & =1.3368
\end{align*}
We can conclude that matrix $\mathbf{A}$ satisfy the restricted isometry property $$0 \le \delta_{K}<1$$ for sparsity $1$, $2$, and $3$. The uniqueness condition require that for sparsity $K$ measurement matrix satisfies restricted isometry property for $2K$ meaning that the uniqueness is guarantied only for $K=1$. For $K=2$ the  condition should be satisfied for $\delta_4$, what is not the case.

The minimization for $K$ sparse vector $\mathbf{X}$ using $\ell_1$-norm will produce the same result as if $\ell_0$-norm were used if  $\delta_{2K}<\sqrt{2}-1.$ It means that there is no guarantee that $\ell_1$ norm minimization could be used in the reconstruction for sparsity $K=1$. Note that different bounds have been derived in literature for this equivalence. One of the derived bounds is that $\delta_{2K}<0.493$. The considered measurement matrix $\mathbf{A}$ would produce a unique solution with $\ell_1$-norm based minimization, according to this bound since $\delta_{2}=0.49<0.493$.  

From this example we can see that uniqueness conditions produce different limits, and that they are very restrictive.

\subsection{Uniqueness of the DFT of Sparse Signals}

In general, the reconstructed signal uniqueness is guarantied if the
restricted isometry property is used and checked. However, two problems exist in the implementation of this approach.
For a specific measurement matrix it produces quite conservative bounds. In
addition, uniqueness check with the restricted isometry property requires a
combinatorial approach, which is an NP hard problem.

In some reconstruction methods the missing measurements are
considered as the minimization variables. The available measurements/samples are
known and fixed. The number of variables in the minimization process is equal
to the number of missing samples/measurements in the observation domain. This
approach is possible when the common signal transforms are the domains of
signal sparsity \cite{Uniq,QRS_2,H_3,H_4,H_5,H_6,SNR_1,HH1}. Then the missing and available samples/measurements form a
complete set of samples/measurements.

The DFT is such a signal sparsity domain. The solution
uniqueness is defined in the sense that the variation of the missing sample
values cannot produce another signal of the same sparsity. In the case when
the signal is already reconstructed then the uniqueness is checked in the
sense that there is no other signal of the same or lower sparsity with the
same set of available samples \cite{Uniq}.
\index{Reconstruction uniqueness}

Consider a signal $x(n)$ with $n\in\mathbb{N}=\{0,1,2,....,N-1\}$.
Assume that $Q$ of its samples at the positions $q_{m}\in\mathbb{N}%
_{Q}=\{q_{1},q_{2},....,q_{Q}\}$ are missing/omitted. The signal is sparse in
the DFT domain, with sparsity $K$. The reconstruction goal is to get $x(n)$,
for all $n\in\mathbb{N}$ using available samples at $n\in\mathbb{M}=\mathbb{N}\backslash
\mathbb{N}_{Q}$\textbf{.} A new signal of the form
\[
x_{a}(n)=x(n)+z(n)
\]
will be analyzed here. For the available signal positions $n\in\mathbb{M}$ the
value of $z(n)$ is fixed  $z(n)=0$, while $z(n)$ may take arbitrary value at
the positions of missing samples $n=q_{m}\in\mathbb {N}_{Q}=\{q_{1}%
,q_{2},....,q_{Q}\}$. If $x(n)$ is a $K$ sparse signal then the DFT of $x_a(n)$ is
\begin{gather*}
X_{a}(k)=X(k)+Z(k)\\
=N\sum_{i=1}^{K}A_{i}\delta(k-k_{0i})+\sum_{m=1}^{Q}z(q_{m})e^{-j2\pi
q_{m}k/N}.
\end{gather*}
Positions of nonzero values in $X(k)$ are $k_{0i}\in\mathbb{K}=\{k_{01}%
,k_{02},....,k_{0K}\}$ with amplitudes $X(k_{0i})=NA_{i}$. The values of
missing samples of $x_{a}(n)=x(n)+z(n)$ for $n\in\mathbb{N}_{Q}$ are
considered as variables. The goal of reconstruction process is to get
$x_{a}(n)=x(n)$, or $z(n)=0$ for all $n\in\mathbb {N}$. This goal should be
achieved by minimizing a sparsity measure of the signal transform $X_{a}(k)$ \cite{Wiley,Ljubisa}.
Existence of the unique solution of this problem depends on the number of
missing samples, their positions, and the signal form.

If a signal with the transform $X(k)$ of sparsity $K$ is obtained using a
reconstruction method, with a set of missing samples, then the reconstruction
$X(k)$ is unique if there is no other signal of the same or lower sparsity
that satisfies the same set of available samples (using the same set of
missing samples as variables).

\textit{Consider a signal }$x(n)$\textit{ that is sparse in
the DFT domain with unknown sparsity. Assume that the signal length is
}$N=2^{r}$\textit{ samples and that }$Q$\textit{ samples are missing at the
instants }$q_{m}\in\mathbb{N}_{Q}$\textit{. Assume that the reconstruction is
performed and that the DFT of reconstructed signal is of sparsity }%
$K$\textit{. The reconstruction result is unique if the inequality }%
\[
K<N-\max_{h=0,1,...,r-1}\left\{  2^{h}\left(  Q_{2^{h}}-1\right)  \right\}
-K
\]
\textit{holds. Integers }$Q_{2^{h}}$\textit{ are calculated as}%
\begin{equation}
Q_{2^{h}}=\max_{b=0,1,...,2^{h}-1}\{\operatorname{card}\{q:q\in\mathbb{N}%
_{Q}\text{ and }\operatorname{mod}(q,2^{h})=b\}\} \label{Qsh}%
\end{equation}

For example, consider a signal with $N=2^{5}=32$ and $Q=9$ missing samples at
\[
q_{m}\in\mathbb{N}_{Q}=\{2,3,8,13,19,22,23,28,30\}.
\]
Using the presented we will find the sparsity limit $K$ when we are able to claim
that the reconstructed sparse signal is unique for any signal form.
\begin{itemize}
\item 
For $h=0$ we use $Q_{2^{0}}=Q$ and get $2^{0}\left(  Q_{2^{0}}-1\right)
-1=(Q-1)-1=9$.
\item 
For $h=1$, {the number }$Q_{2^{1}}${ is the greater value of }%
\[
\operatorname{card}\{q:q\in\mathbb{N}_{Q}\text{ and }\operatorname{mod}%
(q,2)=0\}=\operatorname{card}\{2,8,22,28,30\}=5
\]%
\[
\operatorname{card}\left\{  q:q\in\mathbb{N}_{Q}\text{ and }\operatorname{mod}%
(q,2)=1\right\}  =\operatorname{card}\{3,13,19,23\}=4,
\]
i.e., {the maximal number of even or odd positions of missing samples}. Thus
$Q_{2^{1}}=\max\left\{  5,4\right\}  =5$ with $2^{1}\left(  Q_{2^{1}%
}-1\right)  =8$.
\item
Next $Q_{2^{2}}$ is calculated as the maximal number of missing samples whose
distance is multiple of $4$. For various initial counting positions
$b=0,1,2,3$ the numbers of missing samples with distance being multiple of $4$
are $2,1,3,$ and $3$, respectively. Then $Q_{2^{2}}$ $=\max\left\{
2,1,3,3\right\}  =3$ with $2^{2}(Q_{2^{h}}-1)=8.$
\item
For $Q_{2^{3}}$ the number of missing samples at distances being multiple of
$8$ are found for various $b=0,1,2,3,4,5,6,7$. The value of $Q_{2^{3}}$ is $2$
with $2^{3}(Q_{2^{3}}-1)=8$.
\item
Finally we have two samples at distance $16$ (samples at the positions
$q_{2}=3$ and $q_{5}=q_{2}+N/2$) producing $Q_{2^{4}}=Q_{16}=2$ with
$2^{4}(2-1)=16$.
\end{itemize}

The reconstructed signal of sparsity $K$ is unique if
\begin{gather*}
K<N-\max_{h=0,1,2,3,4}\left\{  2^{h}\left(  Q_{2^{h}}-1\right)  \right\}  -K\\
K<32-\max\left\{  9,8,8,8,16\right\}  -K\\
K<32-16-K
\end{gather*}
or%
\[
K<8.
\]%
An extended discussion about the DFT uniqueness, within the framework of the missing samples as variables, can be found in \cite{Uniq}.

\section{Conclusion}
Sparse signals can be reconstructed from a very reduced set of observations, through compressive sensing. This property has found applications in many fields. The topic of this paper was to introduce the basic definitions in compressive sensing. The conditions for exact and unique reconstruction of original signals are of crucial importance in theory and applications. These conditions are reviewed and related in this paper.   

\section*{Appendix}
MATLAB\textregistered\ functions for spark calculation (Algorithm \ref{Alg1}), restricted isometry constant calculation (Algorithm \ref{Alg2}) and uniqueness test for partial DFT matrix (Algorithm \ref{Alg4}) are provided. Auxiliary function \texttt{nextcomb} used for generation of all possible columns combinations of the measurement matrix used in Algorithms \ref{Alg1} and \ref{Alg2} is given in Algorithm \ref{Alg3}.

%\makeatletter \renewcommand{\ALG@name}{Program} \makeatother

\begin{algorithm}[tbh]
\caption{Measurement matrix spark calculation}
\label{Alg1}
\begin{lstlisting}
function s = spark(A)
% Matrix spark calculation
[M,N] = size(A);
s = M+1;
for k = 1:M
    kk = nchoosek(N,k);
    p = 1:k;
    for m = 1:kk
        A1 = A(:,p);
        if rank(A1) < k
            s = k; break
        end
        p = nextcomb(p,N);
    end
    if s < M+1, break, end
end
\end{lstlisting}
\end{algorithm}

\begin{algorithm}
\caption{Restricted isometry constant calculation}
\label{Alg2}
\begin{lstlisting}
function d = RIP_calc(A,K)
% Restricted isometry constant caluclation
[M,N] = size(A);
d = 0;
kk = nchoosek(N,K);
p = 1:K;
for m = 1:kk
    A1 = A(:,p);
    l = eig(A1'*A1);
    d = max([d, 1-min(l), max(l)-1]);
    p = nextcomb(p,N);
end
\end{lstlisting}
\end{algorithm}

\begin{algorithm}
\caption{Auxiliary function for generation of all combinations}
\label{Alg3}
\begin{lstlisting}
function p = nextcomb(p,N)
% Generate next combination (in lexicographical order)
% Input: previous combination p and number of elements N
% Output: next combination or []

i = length(p);
K = N;
while i > 0 && p(i) == K
    i = i-1;
    K = K-1;
end
if i > 0
    p(i) = p(i)+1;
    for k = i+1:length(p)
        p(k) = p(k-1)+1;
    end
else
    p = [];
end
\end{lstlisting}
\end{algorithm}

\begin{algorithm}[tbh]
\caption{Sparsity limit for partial DFT measurement matrix}
\label{Alg4}
\begin{lstlisting}
function Kt = DFT_check(N,Nq)
% Sparsity limit for partial DFT matrix
% Inputs:
%   N  - total number of samples (must be power of two)
%   Nq - set of missing sample positions

r = log2(N);
if r-round(r) ~= 0, error('N must be power of two'), end
Kt = N;
for h = 0:r-1
    p = rem(Nq,2^h);
    Q = zeros(1,2^h);
    for s = 0:2^h-1
        Q(s+1) = sum(p==s);
    end
    Q2h = max(Q);
    if Kt > (N-2^h*(Q2h-1))/2
        Kt = (N-2^h*(Q2h-1))/2;
    end
end
\end{lstlisting}
\end{algorithm}

\end{document}